\documentclass[secnumarabic,amssymb, nobibnotes, aps, prd]{revtex4-2}
\usepackage{multirow}
\usepackage{graphicx}
\usepackage{caption}
\usepackage{subcaption}

\begin{document}

\title{Maximum mass limit of strange stars in quadratic curvature-matter coupled gravity}%

\author{Debadri Bhattacharjee}%
\email{debadriwork@gmail.com}
\affiliation{IUCAA Centre for Astronomy Research and Development (ICARD), Department of Physics, Cooch Behar Panchanan Barma University, Vivekananda Street, District: Cooch Behar, \\ Pin: 736101, West Bengal, India}
\author{Pradip Kumar Chattopadhyay}
\email{pkc$_$76@rediffmail.com}
\affiliation{IUCAA Centre for Astronomy Research and Development (ICARD), Department of Physics, Cooch Behar Panchanan Barma University, Vivekananda Street, District: Cooch Behar, \\ Pin: 736101, West Bengal, India}
\author{Kazuharu Bamba}
\email{bamba@sss.fukushima-u.ac.jp}
\affiliation{Faculty of Symbiotic Systems Science, Fukushima University, Fukushima 960-1296, Japan}
\begin{abstract}
	We explore the maximum mass limit of strange stars in quadratic curvature gravity with the non-minimal matter coupling. The characteristic parameters of the quadratic curvature coupling and the non-minimal matter coupling imply the contributions from higher-order curvature terms and the coupling between matter and geometry, respectively. We demonstrate, explicitly, that the conservation of energy-momentum tensor can be modified and in the case of negligible non-minimal matter coupling, the formalism of general relativity is recovered. By deriving the Tolman-Oppenheimer-Volkoff equations from the gravitational field equations and applying the MIT bag model equation of state, we obtain the corresponding mass-radius relationships for strange stars. Although the MIT bag model represents a simplified phenomenological equation of state, it remains an effective description of strange quark matter under the extreme conditions prevailing in neutron star/strange star interiors. Within the present framework, the adoption of this equation of state yields stellar radii that are in close agreement with those inferred from recent observations of compact stars as well as GW events. This consistency between theoretical predictions and observational results indicates that, despite its simplicity, the model captures essential features of dense matter and supports the reliability of the results reported in this work. Furthermore, we show that the maximum mass limit of strange stars can exceed the general relativistic counterpart. Specifically, we find that a maximum mass up to 3.11 solar mass is achievable which suggests that the lighter companion of GW190814 could plausibly be a strange star.  
\end{abstract}
\maketitle
	   
\section{Introduction}\label{sec1} 
The complex internal composition of a neutron star (NS) has been a long-standing puzzle for astrophysicists for decades. To date, it has been established that the correct approach to describe a NS relies on the choice of Equation of State (EoS) \cite{Lattimer}. For a particular EoS, the numerical solution of the Tolman-Oppenheimer-Volkoff equations \cite{Tolman,Oppenheimer} can yield the maximum mass and the associated radius of any NS configuration. Moreover, the maximum mass-radius and the characterisation of the internal matter distribution through an EoS provide a possible avenue to study the evolutionary process of the progenitor as well as the plausible process responsible for the NS formation. Further, the core density of NS reaches the order of $10^{14}-10^{15}~\mathrm{gm/cm^{3}}$ and in such a high-density environment, the physical nature of the internal matter is subjected to extreme conditions of gravity. Given the current limitations in replicating such strong gravity regimes in laboratory settings, we must often depend on theoretical models to explore these phenomena. Although numerous EoSs are available, their reliability depends critically on astrophysical measurements obtained through various observational probes. These measurements are vital for determining the macroscopic properties of NSs as well as assessing the softness or stiffness of the pressure-density relation, which in turn helps constrain a realistic EoS \cite{Hebeler,Ozel1,Steiner}. If the mass-radius relationship predicted through a theoretical choice of EoS aligns well with observational results, it can significantly narrow down the choice of viable EoS. This, in turn, allows for a more reliable investigation of the internal structure of NSs under strong gravitational fields. 

Although there have been several attempts to model the interior of NS using different EoS, the proper explanation of physical properties of many stars (mass, radius, compactness etc.), estimated from recent observations, is not possible. The quark star hypothesis, where the compact star is hypothesised to be composed of Strange Quark Matter (SQM), has emerged as a prime candidate to describe the internal matter composition of such stars on the basis of observational results. Under the assumption of zero external pressure, Madsen \cite{Madsen} demonstrated that quark stars, which are composed solely of $u$ and $d$ quarks, are inherently unstable. However, the incorporation of $s$ quarks into the system lowers the energy per baryon significantly compared to the two-flavour configuration. This, in turn, enhances the overall stability of the system. Hence, the presence of $s$ quarks is necessary for a stable quark matter configuration. Consequently, compact stars composed of $u$, $d$, and $s$ quarks are classified as Strange Stars (SS) or Strange Quark Stars (SQS) \cite{Alcock,Madsen}. Witten \cite{Witten} further postulated that Strange Quark Matter (SQM), comprising equal numbers of $u$, $d$, and $s$ quarks, approximately, may represent the true ground state of Quantum Chromodynamics (QCD), in the condition that its energy per baryon is lower than that of the most stable atomic nucleus, $^{56}$Fe. This hypothesis lends strong theoretical support to the absolute stability of SQM. In the high-energy limit, QCD predicts asymptotic freedom, wherein nucleons dissociate into their fundamental constituents while forming a weakly interacting quark-gluon plasma (QGP) phase. Alford \cite{Alford} proposed that the conditions in the dense and relatively cold cores of NS may be sufficient to trigger the formation of de-confined quark matter. Although the transition from hadronic matter to quark matter is not yet fully understood, various phenomenological approaches have been developed to model such process, {\it viz.}, MIT bag model EoS \cite{Chodos}, Nambu-Jona-Lasinio (NJL) model \cite{NJL}, Colour-Flavour-Locked (CFL) phase \cite{Lugones1}, Hyperonic EoS \cite{Wu}, kaon-meson condensate including hyperons \cite{Ma}, etc. Among these models, the MIT bag model EoS \cite{Chodos} has been one of the prime candidates to describe the interior of SQS, through the thermodynamic behaviour of de-confined quark matter within a finite region, or `bag'. This bag represents the energy difference between the perturbative and non-perturbative QCD vacuum. Further, in MIT bag model, the quarks are considered to be degenerate Fermi gas, composed of $u$, $d$, $s$ quarks and electrons. Though simplified MIT bag model EoS does not cover all microphysical complexities of SQM, it may be useful to study the properties of SQM which are in agreement with the results obtained from recent observations. Several studies \cite{Brilenkov,Paulucci,Arbanil,Lugones} have been conducted to explain the properties of SS using the MIT bag model EoS.

The correct extraction of the mass limit is only possible through observational evidence pertaining to the NS binary systems. Particularly, in the case of radio pulsars, the most accurate mass measurement provides a mass value of $1.35~\mathrm{M_{\odot}}$ \cite{Thorsett}. In recent years, a significant volume of observational data from pulsars and gravitational wave (GW) events has been collected and analysed by various researchers. These analyses have led to precise estimations of several physical properties of compact astrophysical objects. Notably, using the Shapiro delay technique, the mass of the millisecond pulsar PSR J0740+6620 was measured to be $(2.14)^{+0.10}_{-0.09}~\mathrm{M{\odot}}$ \cite{Cromartie}, representing one of the most massive known pulsar. This measurement has intensified discussions regarding the upper mass limit of compact stars. Moreover, it has been suggested that certain compact objects in interacting binary systems could possess masses exceeding that of PSR J0740+6620 \cite{Cromartie}. In particular, the gravitational wave event GW190814 revealed a secondary component with a mass in the range of $2.5$-$2.67~\mathrm{M{\odot}}$ at 90\% confidence \cite{Abbott}. However, the exact nature of this object remains uncertain, it could either be an unusually massive NS, a SS or a low-mass black hole. If the former scenario holds true, it would necessitate theoretical advancements to extend the maximum mass to such high values. Developing such a framework is also crucial for probing the internal composition of matter at densities beyond nuclear saturation.

However, achieving such high mass values and still maintaining the viability conditions in the context of General Relativity (GR) seems unlikely. Furthermore, the exact upper mass bound of NS/SS is still elusive. Step by step, the latest observational results show that we are in ample need of new theory that can describe the recent astrophysical observations with greater accuracy. As a result, we turn to the modified gravity approach. Modified gravity models, especially $f(R)$ \cite{Sotiriou,DeFelice}, $f(T)$ \cite{Cai}, $f(Q)$ \cite{Heisenberg}, $f(G)$ \cite{Nojiri1}, extended theories of gravity \cite{Capozziello1,Clifton,Nojiri1} etc. have demonstrated improved fits for known issues like cosmic acceleration \cite{Sahni}, dark energy, and dark matter \cite{Joyce,Peebles,Paddy,Paddy1,Durrer}, and have been applied successfully to determine upper mass limits for compact stars \cite{Astashenok1,Astashenok2,Astashenok3,Astashenok4,Astashenok5,Astashenok6,Astashenok7,Astashenok8}. Among these, the $f(R)$ gravity model offers one of the most straightforward generalisations, in which the conventional Einstein-Hilbert action is substituted by a generic function of the Ricci scalar $R$ \cite{Nojiri}. The $f(R)$ framework has provided significant insights into galactic kinematics and the accelerated expansion of the universe \cite{Capozziello, Borowiec}. Building on this, Harko et al. \cite{Harko} extended the action further by incorporating the trace of the energy-momentum tensor $(T)$, leading to the $f(R,T)$ gravity theory. In their pioneering work, they interpreted the $T$ dependence as a potential signature of quantum effects or exotic imperfect fluids \cite{Harko1}. It revealed distinctive features such as a non-vanishing covariant derivative of the energy-momentum tensor along with an additional acceleration component arising from matter-curvature coupling. 

Despite these innovations, $f(R,T)$ gravity presents several critical problems. Among them, the most prominent are the ambiguities in the formulation, potential violations of energy-momentum conservation, model-dependent instabilities, and challenges in matching observational constraints. Direct coupling of the energy-momentum tensor to curvature often complicates the field equations. It can result in unconventional matter dynamics or ambiguities in interpreting the physical source terms. Further, linear couplings to $T$ may produce models that fail to account for the observed cosmic acceleration. To address these limitations, a possible outlook may be the extensions involving higher-order curvature invariants and non-minimal curvature-matter coupling. These additions enhance the flexibility of the model and allow better fits to cosmological data, while enabling transitions between deceleration and acceleration in cosmic expansion scenarios. Moreover, such generalised theories can stabilise solutions like de Sitter and power-law cosmologies, restore compatibility with energy conditions, and help resolve non-conservation issues by realigning the gravitational and matter sectors.

The motivation for considering the modified gravity model, $f(R,T)=R+\alpha R^{2}+2\beta T$ gravity model in the context of compact stars is twofold. First, compact objects such as NS and SS provide unique laboratory for testing gravity in the strong-field and high-density regime, where deviations from GR may naturally emerge and cannot be constrained solely by solar-system or cosmological observations. In particular, quadratic curvature corrections and explicit gravity-matter coupling are expected to become relevant at supranuclear densities, potentially modifying the internal structure and global properties of compact stars. Second, recent observations of massive compact objects, together with radius measurements and gravitational-wave constraints, have highlighted persistent tensions between theoretical predictions based on GR and conventional EoS, often requiring the introduction of exotic matter or extreme stiffness of the EoS. Within this context, the present model offers a minimal yet physically motivated extension of GR that allows one to explore whether such tensions can be alleviated through modified gravity effects alone. The inclusion of the $R^{2}$ term captures higher-order curvature contributions that are well motivated by both quantum gravity considerations and successful inflationary models, while the linear $T$ dependence introduces an effective gravity-matter interaction that can significantly influence the equilibrium configuration of ultra-dense matter. By studying SS described by the MIT bag model EoS, this work aims to study the impact of modified gravity on the mass-radius relation and the maximum mass limit, without invoking additional exotic components. Ultimately, these advanced frameworks offer more robust tools for modeling the strong-field regime and reconciling theoretical predictions with the latest astrophysical observations.

In this paper, we present a novel approach to redefine the maximum mass limit of SS, by considering the higher-order curvature correction along with the gravity-matter coupling. This theory is effectively expressed as, $f(R,T)=R+\alpha R^{2}+2\beta T$ gravity, where $R$ and $T$ denote the Ricci scalar and trace of energy momentum tensor, respectively. Additionally, $\alpha$ and $\beta$ characterise the strength of higher-order curvature correction and gravity-matter coupling. It must be noted that $f(R+\alpha R^{2})$ gravity is renowned for Starobinsky inflation \cite{Starobinsky}, consistent with Planck 2018 data in cosmology \cite{Akrami}. This model is a more general framework of the modified gravity theories in comparison to the results obtained in Refs. \cite{Feola,Capozziello2}. It is important to note that in the high-density, strong-gravity environment within NS/SS, higher-order curvature effects may arise naturally \cite{Astashenok,Fiziev,Stoykov1,Stoykov2}. Furthermore, to investigate the behaviour of matter under such extreme conditions and its influence on the EoS, and the resulting maximum mass limit, we have developed this new framework of an extended theory of gravity. Additionally, the new effective parameters arising from this theory can, potentially, remove some of the inconsistencies in the theoretical formalism of compact stars that are, in general, addressed through the use of exotic matter distribution. The main objective of this study is to incorporate quadratic curvature corrections and gravity-matter coupling into the TOV framework to derive the mass-radius relationship using the MIT bag model EoS. Despite its phenomenological and simplified nature, the MIT bag model EoS provides a widely used description of strange quark matter in ultra-dense astrophysical environments. In this work, we employ this EoS to explore the possibility whether the present model can yield physically meaningful predictions for compact stars (especially SS), particularly with respect to stellar radii. The aim is to examine the extent to which the model outcomes can be compared with the recent observational data on compact stars as well as the secondary objects in many GW events. Furthermore, by systematically varying the parameters $\alpha$ and $\beta$, we aim to investigate the deviations from GR predictions, with a particular focus on identifying the emergence of strong gravity effects and their influence on the maximum mass limit of SS.  

In the present framework, we couple the matter sector to the higher-order curvature gravity. A caveat is in order regarding the interpretation of the $2\beta T$ term. Although $T$ is a scalar built from matter, a term linear in $T$ in the gravitational action is not equivalent to the introduction of an independent propagating scalar field. The $R^{2}$ term produces a genuine scalar degree of freedom (the scalaron) after a Legendre/conformal transformation because it leads to a kinetic structure for that mode. In contrast, the $2\beta T$ term acts as a direct algebraic coupling between curvature and the matter trace and contains no kinetic term. Thus it modifies the geometry-matter coupling, and generally leads to $\nabla^\mu T_{\mu\nu}\neq0$, but does not introduce a new propagating scalar by itself. If one desires a scalar-tensor representation where the matter trace is carried by an independent scalar field, that scalar must be introduced explicitly, for example, by promoting $T$ to a field $\chi$ with a kinetic term and potential. In this work we maintain the phenomenological $2\beta T$ coupling and discussed its physical consequences for stellar structure. We also examine limiting cases, and the dependence on the chosen matter Lagrangian. Previously, the studies of $f(R)$, or $f(R+\alpha R^{2})$ showed that in the high energy regime of compact stars, the inclusion of higher-order curvature provides a viable theoretical construct to study high mass compact stars. However, in presence of such high energy environment, the matter sector is bound to be influenced, and accounting for the matter contribution gives a complete foundation to study such extreme conditions of gravity. Keeping this in mind, we construct the present framework. 

The main contents of the paper are organised as follows: Section~\ref{sec2} describes the mathematical formalism in the framework of $f(R+\alpha R^{2},T)$ gravity. Here, we develop the modified field equations for a static, spherically symmetric space-time and isotropic interior matter distribution. Section~\ref{sec3} deals with the formulation of TOV equations in this new context. The mass and pressure gradients are obtained from the modified field equations. The main results concerning the maximum mass and the associated radius are illustrated and tabulated in Section~\ref{sec4}. In this section, we obtain the particular ranges for the parameters that yield physically acceptable TOV solutions. Further, we study the stability of the model through the evaluation of adiabatic index and Harrison-Zel'dovich-Novikov condition in Section~\ref{sec5}. The possible tests that may lead to strange star detection, thereby validating the present framework, are addressed in Section~\ref{sec6}. Finally, we summarise the key features, originating from this study, in Section~\ref{sec7}.
\section{Formalism of quadratic curvature-matter coupled gravity}\label{sec2} In the metric formalism, the Einstein-Hilbert action takes the form:
\begin{equation}
	\mathcal{S}=\frac{1}{16\pi}\int{\sqrt{-g}f(\tilde{R},T)d^{4}x}+\int{\sqrt{-g}\mathcal{L_{M}}d^{4}x}, \label{eq1}
\end{equation}
where we have considered the system of units, $G=1$ and $c=1$. In Eq.~(\ref{eq1}), the modified Ricci scalar is defined as, $\tilde{R}=R+\alpha R^{2}$, where $\alpha$ is the curvature correction parameter. In the present analysis, we consider the following form: $f(\tilde{R},T)=R+\alpha R^{2}+2\beta T$, which explicitly includes the quadratic curvature modifications along with the gravity-matter coupling, through the trace of energy-momentum tensor $T$, governed by the coupling parameter $\beta$. Moreover, $\mathcal{L_{M}}$ is the Lagrangian associated with the matter sector. Notably, this form of $f(\tilde{R},T)$ provides the following advantages and motivations:
\begin{itemize}
	\item The $R^{2}$ correction arises naturally in semi-classical gravity as a one-loop quantum effect, and it supports a successful high-curvature inflationary phase (Starobinsky inflation) \cite{Starobinsky}.
	\item In the cosmological aspect, these type of models may dynamically support self-consistent de Sitter-like solutions for large $R$.  
	\item In such matter-gravity coupled theories, the viability conditions $f_{R}>0$ and $f_{RR}>0$ demand $\alpha>0$, which will avoid ghost modes and the Dolgov-Kawasaki instability \cite{Dolgov}.
	\item In such extended $f(R,T)$ model, a small $\alpha$ yields a sufficiently massive scalaron, recovering GR behaviour in the low-curvature regime and satisfying Solar-System bounds.
	\item The linear $T$-dependence enables matter-geometry coupling governed by $\beta$, offering phenomenological flexibility in strong-gravity environments such as compact stars.
	\item This model may predict stable, causal interior solutions with realistic EoS when $\alpha>0$ and $|\beta|$ is small. 
	\item The theory smoothly recovers GR in the limit $\alpha\rightarrow0$ and $\beta\rightarrow0$, indicating that the present framework is a minimal and well-motivated modification of Einstein gravity.
\end{itemize}
Now, taking the variation of Eq.~(\ref{eq1}) with respect to the fundamental metric tensor $g_{ij}$, we obtain:
\begin{equation}
	f_{R}R_{ij}-\frac{1}{2}g_{ij}f(\tilde{R},T)+f_{R}\Big[g_{ij}\Box-\nabla_{i}\nabla_{j}\Big]=8\pi T_{ij}-f_{T}T_{ij}-f_{T}\Theta_{ij}. \label{eq2}
\end{equation}
Here, $f_{R}=\frac{df(\tilde{R},T)}{dR}$, $f_{T}=\frac{df(\tilde{R},T)}{dT}$, $\Box\equiv~\frac{\partial_{i}(\sqrt{-g}g^{ij}\partial_{j})}{\sqrt{-g}}$ represents the d'Alembert operator, $\nabla_{i}$ denotes the covariant derivative associated with the Levi-Civita connection corresponding to the metric $g_{ij}$, and $\Theta_{ij}=g^{ij}\Big(\frac{\delta T_{ij}}{\delta g_{ij}}\Big)$ representing the variation of the energy-momentum tensor with respect to the metric. Now considering the form, $f(\tilde{R},T)=R+\alpha R^{2}+2\beta T$ and refining Eq.~(\ref{eq2}), we obtain:
\begin{eqnarray}
	G_{ij}+2\alpha R R_{ij}-\frac{1}{2}g_{ij}\alpha R^{2}+2\alpha\Big[g_{ij}\Box-\nabla_{i}\nabla_{j}\Big]R\nonumber\\=8\pi T_{ij}-f_{T}T_{ij}-f_{T}\Theta_{ij}+g_{ij}\beta T, \label{eq3}
\end{eqnarray}
where $R_{ij}$ is the Ricci tensor and $G_{ij}$ represents the Einstein tensor. Notably, for $\alpha=0$ and $\beta\neq0$, we obtain the $f(R,T)$ formalism presented by Harko et al. \cite{Harko}, whereas, for $\beta=0$ and $\alpha\neq0$, we obtain the $f(R)$ formalism \cite{Buchdahl,Bertolami}. 

To describe a compact star in a static and spherically symmetric space-time, we start with the line element of the form:
\begin{equation}
	ds^2=-e^{2\nu(r)}dt^2+e^{2\lambda(r)}dr^2+r^2(d\theta^2+sin^2\theta d\phi^2), \label{eq4}
\end{equation}
where $\lambda(r)$ and $\nu(r)$ are metric potentials that depend on the radial coordinate only. Moreover, considering $\mathcal{L_{M}}=p$, we obtain the isotropic interior matter distribution in the following form:
\begin{equation}
	T_{ij}=(p+\rho)u_{i}u_{j}+pg_{ij}, \label{eq5}
\end{equation} 
where $p$ denotes the isotropic pressure, $\rho$ represents the energy density, and $u^{i}u_{i}=-1$ specifies the normalisation condition for the four-velocity vector. It follows from Eqs.~(\ref{eq4}) and (\ref{eq5}) that $T=-\rho+3p$, $\Theta_{ij}=-2T_{ij}+pg_{ij}$ and $\Theta=-2T+4p$. Now, the non-conservation of the energy-momentum tensor leads to the form:
\begin{eqnarray}	
	\nabla^{i}T_{ij}=\frac{f_{T}(\tilde{R},T)}{8\pi-f_{T}(\tilde{R},T)}\Big[(T_{ij}+\Theta_{ij})\nabla^{i}~ln~f_{T}(\tilde{R},T)\\\nonumber+\nabla^{i}\Theta_{ij}-\frac{1}{2}g_{ij}\nabla^{i}T\Big]. \label{eq6}
\end{eqnarray}
Moreover, by using the present form $f(\tilde{R},T)=R+\alpha R^{2}+2\beta T$, we find that Eq.~(\ref{eq6}) takes the form:
\begin{equation}
	\nabla^{i}T_{ij}=\frac{2\beta}{8\pi+2\beta}\Big[\nabla^{i}(pg_{ij}-\frac{1}{2}g_{ij}\nabla^{i}T)\Big]. \label{eq7}
\end{equation}
Interestingly, Eq.~(\ref{eq7}) is similar to the form obtained by Pretel et al. \cite{Pretel} in the context of $f(R,T)$ gravity. Additionally, for $\beta\rightarrow0$, we retain the formalism of GR.  

\section{Obtaining the Tolman-Oppenheimer-Volkoff equations}\label{sec3} 
Considering the static and spherically symmetric line element described in Eq.~(\ref{eq4}) and the isotropic perfect fluid distribution expressed in Eq.~(\ref{eq5}), the modified $G_{tt}$ and $G_{rr}$ components of the field Eq.~(\ref{eq3}) are obtained as:
\begin{eqnarray}
	\frac{2\lambda'e^{-2\lambda}}{r}+\frac{1-e^{-2\lambda}}{r^{2}}+2\alpha R\Big[e^{-2\lambda}\Big\{-\nu''-\nu'^{2}+\lambda'\nu'-\frac{2\nu'}{r}\Big\}\Big]\nonumber\\+\frac{1}{2}\alpha R^{2}-2\alpha\Big[e^{-2\lambda}\Big\{R''+(\nu'-\lambda'+\frac{2}{r})R'\Big\}\Big]=8\pi\rho\nonumber\\+3\beta\rho-\beta p,\nonumber\\ \label{eq8}
\end{eqnarray}
and
\begin{eqnarray}
	\frac{2\nu'e^{-2\lambda}}{r}-\frac{1-e^{-2\lambda}}{r^{2}}+2\alpha R\Big[e^{-2\lambda}\Big\{\nu''+\nu'^{2}-\lambda'\nu'-\frac{2\lambda'}{r}\Big\}\Big]\nonumber\\-\frac{1}{2}\alpha R^{2}+2\alpha\Big[e^{-2\lambda}\Big\{R''+(\nu'-\lambda'+\frac{2}{r})R'\Big\}\Big]-2\alpha\Big[R''-\lambda'R'\Big]\nonumber\\=8\pi p+3p\beta-\beta\rho.\nonumber\\ \label{eq9}
\end{eqnarray}
Now, using the explicit result, $e^{-2\lambda}=1-\frac{2m(r)}{r}$ and the present form of $f(\tilde{R},T)$ in Eqs.~(\ref{eq7}), (\ref{eq8}) and (\ref{eq9}), we obtain the modified TOV equation \cite{Tolman,Oppenheimer} in the form:
\begin{eqnarray}
	\frac{dm}{dr}=4\pi r^{2}\rho+\frac{\beta r^{2}}{2}\Big[3\rho-p\Big]+\frac{r^{2}}{2}\Bigg[-2\alpha R\Big[e^{-2\lambda}\Big\{-\nu''-\nu'^{2}\nonumber\\+\lambda'\nu'-\frac{2\nu'}{r}\Big\}\Big]-\frac{1}{2}\alpha R^{2}+2\alpha\Big[e^{-2\lambda}\Big\{R''+(\nu'-\lambda'+\frac{2}{r})R'\Big\}\Big] \Bigg], \nonumber\\\label{eq10}
\end{eqnarray}
and 
\begin{equation}
	\frac{dp}{dr}=\Bigg[\frac{\beta}{8\pi+3\beta}\Bigg]\rho'-\Bigg[\frac{8\pi+2\beta}{8\pi+3\beta}\Bigg](\rho+p)\Bigg[\frac{m(r)+4\pi r^{3}p_{eff}}{r[r-2m(r)]}\Bigg], \label{eq11}
\end{equation}
where, $p_{eff}=p+\frac{\beta}{8\pi}\Big[3p-\rho\Big]-\frac{1}{8\pi}\Bigg[2\alpha RR_{r}^{r}-\frac{1}{2}\alpha R^{2}+2\alpha\Box R-2\alpha\nabla_{r}\nabla_{r}R\Bigg]$ and the prime $(')$ indicates derivatives with respect to $r$. To obtain the maximum mass-radius relation, we simultaneously solve Eqs.~(\ref{eq10}) and (\ref{eq11}) for a particular choice of EoS, using the boundary conditions, $m(0)=0$ and $\rho(0)=\rho_{c}$. 

\section{Maximum mass-radius using MIT Bag model EoS}\label{sec4} The MIT bag model \cite{Chodos} is one of the most successful theories for describing the interior of SS. Over time, several studies have been conducted that centred around the MIT bag model. Initially, the EoS contained a constant bag parameter $(B_{g})$ \cite{Kettner}. According to the predictions made by Madsen \cite{Madsen}, the allowed range of the bag constant $B_{g}$ is constrained by the requirement of strange matter stability. The lower bound is set at $B_{g}^{1/4}=145~\mathrm{MeV}$, corresponding to $B_{g}=57.55~\mathrm{MeV/fm^3}$, below which non-interacting strange quark matter becomes unstable. In the scenario involving only $u$ and $d$ quarks, the system may not be stable energetically, as indicated in Ref. \cite{Madsen}. The inclusion of the $s$-quark lowers the energy per baryon, thereby enhancing the stability of the system. For two-flavour (non-strange) quark matter to be stable relative to nuclear matter, its energy per baryon must remain below the neutron mass, $939.6~\mathrm{MeV}$ \cite{Madsen} at zero external pressure. This condition sets the minimum value of $B_{g}$ necessary to prevent atomic nuclei from decaying into non-strange quark matter. On the other hand, the upper bound on $B_{g}$ arises from the requirement that strange quark matter be stable compared to iron nuclei. This bound corresponds to $(B_{g})^{1/4}_\mathrm{{max}}= 162.8~\mathrm{MeV}$ or equivalently $(B_{g})_\mathrm{{max}}\\=91.54~\mathrm{MeV/fm^3}$ \cite{Madsen}. When considering stability relative to a neutron under zero external pressure, the upper limit shifts slightly to $(B_{g})^{1/4}_\mathrm{{max}}= 164.4~\mathrm{MeV}$, yielding 
$(B_{g})_\mathrm{{max}}=95.11~\mathrm{MeV/fm^3}$ \cite{Kapusta,Madsen}. 

In this section, we have solved the modified TOV equations using the MIT EoS for the upper and lower bounds of the bag parameter, $B_{g}$. Notably, the choice of $\alpha$ and $\beta$ is crucial here. In higher-order curvature-matter coupled gravity, the appearance of ghost field is a problematic scenario. To address this situation, we have ensured a positive kinetic term by imposing the condition $f_{RR}\geq0$. In the present model, $f_{RR}\geq0$ leads to $\alpha>0$. Further, following the previous studies \cite{Deb,Carvalho}, we have noted that $\beta$ can take both positive and negative values. Hence, we have considered systematic variations in $\alpha$ and $\beta$ to obtain the maximum mass-radius relation. Notably, the ranges of $\alpha$ and $\beta$ are chosen carefully so that the solution of the TOV equation remains well-behaved. Based on this setup, the resulting findings are presented and discussed below: 
\begin{itemize}
	\item {\bf Case-I:} We have taken $B_{g}=57.55~\mathrm{MeV/fm^3}$, and have considered the range of $\beta$ from $-0.5$ to $0.5$. Now, $\alpha$ is varied within a suitable range to obtain the mass-radius plot for SS, and the results are illustrated in Figs.~\ref{fig1}, \ref{fig2}, \ref{fig3}, and \ref{fig4}, respectively. 
	\begin{figure*}[t!]
		\centering
		\begin{minipage}{0.45\linewidth}
			\hspace{-0.5cm}
			\includegraphics[width=8.5cm]{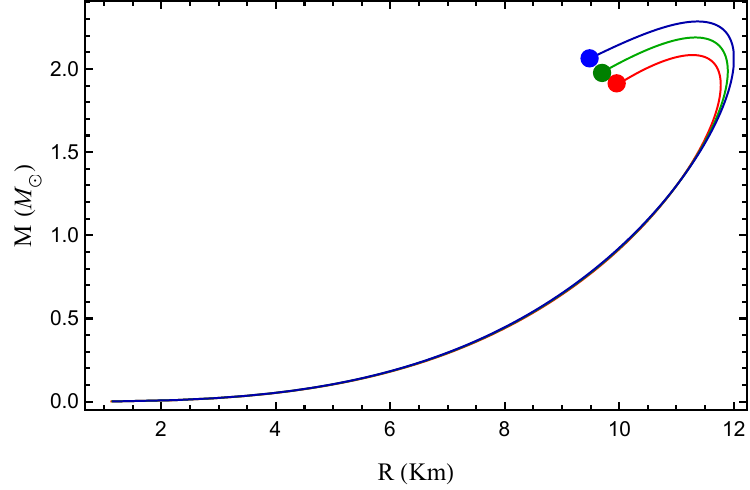}
			\subcaption{}
			\label{fig1}
		\end{minipage}
		\hfill
		\begin{minipage}{0.45\linewidth}
			\hspace{-0.4cm}
			\includegraphics[width=8.5cm]{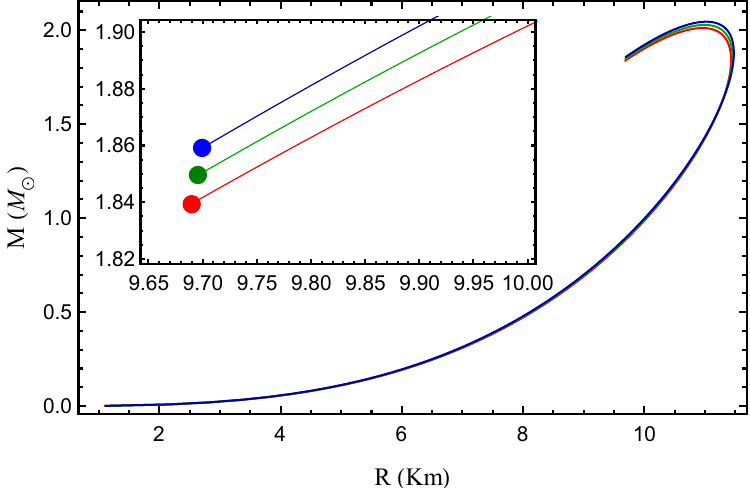}
			\subcaption{}
			\label{fig2}
		\end{minipage}
		\begin{minipage}{0.45\linewidth}
			\hspace{-0.5cm}
			\includegraphics[width=8.5cm]{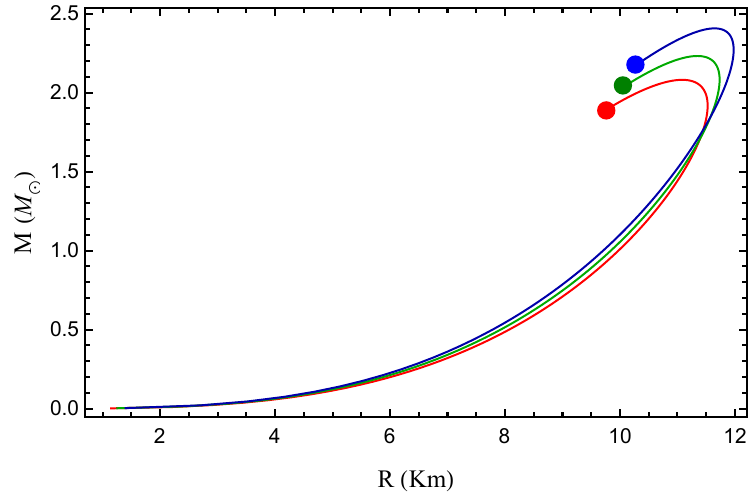}
			\subcaption{}
			\label{fig3}
		\end{minipage}
		\hfill
		\begin{minipage}{0.45\linewidth}
			\hspace{-0.5cm}
			\includegraphics[width=8.5cm]{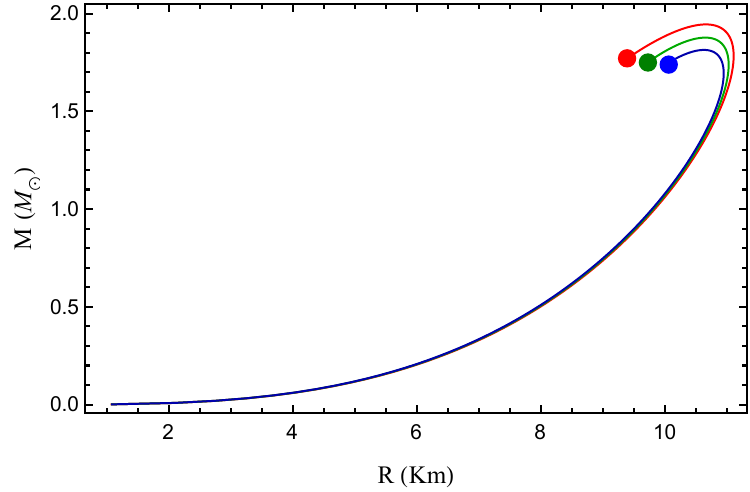}
			\subcaption{}
			\label{fig4}
		\end{minipage}
		\caption{Mass-radius relation for $B_{g}=57.55~\mathrm{MeV/fm^3}$, and different combinations of $\alpha$ and $\beta$. Fig.~\ref{fig1}.: $\beta=-0.5$ with $\alpha=0,~2.5,~4.7$ (red, green, blue). Fig.~\ref{fig2}.: $\beta=0$ with $\alpha=0,~2.5,~5$ (red, green, blue). Fig.~\ref{fig3}.: $\beta=0$ with  $\alpha=10,~30,~50$ (red, green, blue). Fig.~\ref{fig4}.: $\beta=0.5$ with $\alpha=0,~2.5,~5$ (red, green, blue).}
		\label{fig1a}
	\end{figure*}
	The maximum mass and the associated radius corresponding to the Figs.~\ref{fig1}, \ref{fig2}, \ref{fig3}, and \ref{fig4} are tabulated in Tables~\ref{tab1}, \ref{tab2}, \ref{tab3}, and \ref{tab4}, respectively.
	\begin{table*}[t!]
		\centering
		\caption{Maximum mass and radius for $B_{g}=57.55~\mathrm{MeV/fm^3}$ with different $\alpha$ and $\beta$. Panels (a)--(d) correspond to Figs.~\ref{fig1}--\ref{fig4}.}
		\begin{subtable}[t]{0.45\textwidth}
			\centering
			\caption{$\beta=-0.5$}
			\label{tab1}
			\begin{tabular}{ccc}\hline
				$\alpha$ & $M_{\rm max}(M_{\odot})$ & $R$ (Km) \\ \hline
				0   & 2.08  & 11.28 \\
				2.5 & 2.19  & 11.33 \\
				4.7 & 2.28  & 11.37 \\\hline
			\end{tabular}
		\end{subtable}
		\hspace{1em}
		\begin{subtable}[t]{0.45\textwidth}
			\centering
			\caption{$\beta=0$}
			\label{tab2}
			\begin{tabular}{ccc}\hline
				$\alpha$ & $M_{\rm max}(M_{\odot})$ & $R$ (Km) \\ \hline
				0   & 2.012 & 10.96 \\
				2.5 & 2.030 & 11.00 \\
				5   & 2.046 & 11.02 \\\hline
			\end{tabular}
		\end{subtable}
		\vspace{1em}
		\begin{subtable}[t]{0.45\textwidth}
			\centering
			\vspace{0.2cm}
			\caption{$\beta=0$ (large $\alpha$)}
			\label{tab3}
			\begin{tabular}{ccc}\hline
				$\alpha$ & $M_{\rm max}(M_{\odot})$ & $R$ (Km) \\ \hline
				10  & 2.08 & 11.08 \\
				30  & 2.23 & 11.35 \\
				50  & 2.41 & 11.64 \\\hline
			\end{tabular}
		\end{subtable}
		\hspace{1em}
		\begin{subtable}[t]{0.45\textwidth}
			\centering
			\vspace{0.2cm}
			\caption{$\beta=0.5$}
			\label{tab4}
			\begin{tabular}{ccc}\hline
				$\alpha$ & $M_{\rm max}(M_{\odot})$ & $R$ (Km) \\ \hline
				0   & 1.94 & 10.67 \\
				2.5 & 1.87 & 10.66 \\
				5   & 1.81 & 10.64 \\ \hline
			\end{tabular}
		\end{subtable}
	\end{table*}
	Based on these results, we infer the following:
	\begin{itemize}
		\item From Fig.~\ref{fig1}, we note that for a negative choice of $\beta$, the maximum mass-radius increases with increasing $\alpha$. This may be related to the fact that negative gravity-matter coupling indicates a weakly coupled system that allows for a larger configuration. Further, with increasing curvature corrections, the stellar structure may counter gravity more strongly and withstand the gravitational collapse. Consequently, the mass and the associated radius increase. However, from Table~\ref{tab1}, we have noted that for $\beta=-0.5$, we can extract the maximum mass and radius up to $\alpha=4.7$, beyond which TOV equations cannot be solved properly.
		\item In GR, solving the TOV equations with the MIT bag EoS yields a maximum mass of $2.012~\mathrm{M_{\odot}}$. Setting $\alpha=0$ and $\beta=0$ reproduces this result as shown in Table~\ref{tab2}, confirming that the formalism reduces to GR in this limit.
		\item In case of $\beta=0$, the upper limit of $\alpha$ increases to $\alpha=52$ which yields a maximum mass of $2.42~\mathrm{M_{\odot}}$ and a radius of $11.68~\mathrm{Km}$. Beyond $\alpha=52$, the solution of TOV equations does not produce acceptable results. 
		\item Figure~\ref{fig4} and the corresponding Table~\ref{tab4}, show that for positive value of $\beta$, i.e., $\beta=0.5$, the maximum mass decreases with increasing curvature corrections. This may be attributed to the fact that the strength of gravity-matter coupling increases with positive $\beta$. Further, the inclusion of increasingly higher-order curvature corrections modifies the effective field equations. These two effects may lead to a softer EoS for the stellar structure. As a result, the maximum mass decreases. Additionally, we note that for $\beta=0.5$, the maximum permissible value of $\alpha$ reaches $\alpha=10$ and at this upper bound, the maximum mass and radius are obtained as $1.70~\mathrm{M_{\odot}}$ and $10.55~\mathrm{Km}$.	
	\end{itemize}   	
\end{itemize}
\begin{itemize}
	\item {\bf Case-II:} Here, we consider the upper bound of bag parameter, $B_{g}=95.11~\mathrm{MeV/fm^3}$ with similar range of $\beta$ and appropriate variation of $\alpha$ to obtain the mass-radius relation, and the results are demonstrated in Figs.~\ref{fig5}, \ref{fig6}, \ref{fig7}, and \ref{fig8}, respectively. 
	\begin{figure*}[t!]
		\centering
		\begin{minipage}{0.45\linewidth}
			\hspace{-0.5cm}
			\includegraphics[width=8.5cm]{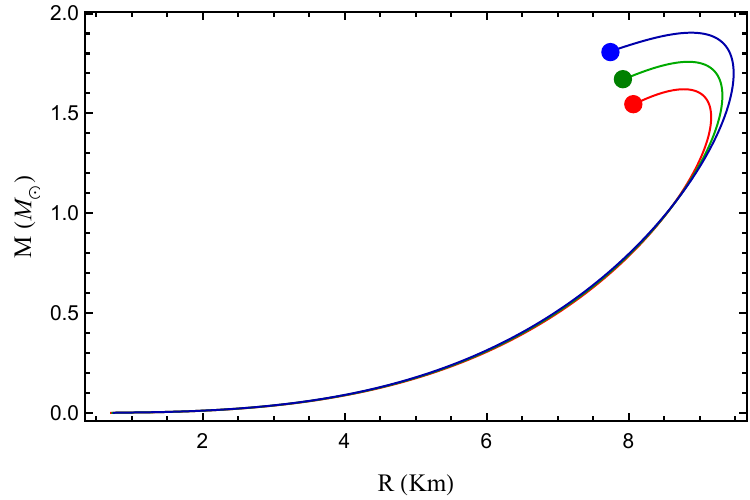}
			\subcaption{}
			\label{fig5}
		\end{minipage}
		\hfill
		\begin{minipage}{0.45\linewidth}
			\hspace{-0.4cm}
			\includegraphics[width=8.5cm]{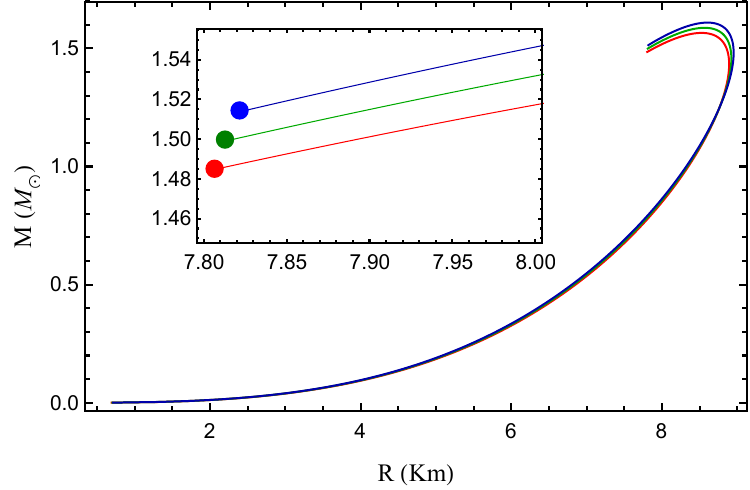}
			\subcaption{}
			\label{fig6}
		\end{minipage}
		\begin{minipage}{0.45\linewidth}
			\hspace{-0.5cm}
			\includegraphics[width=8.5cm]{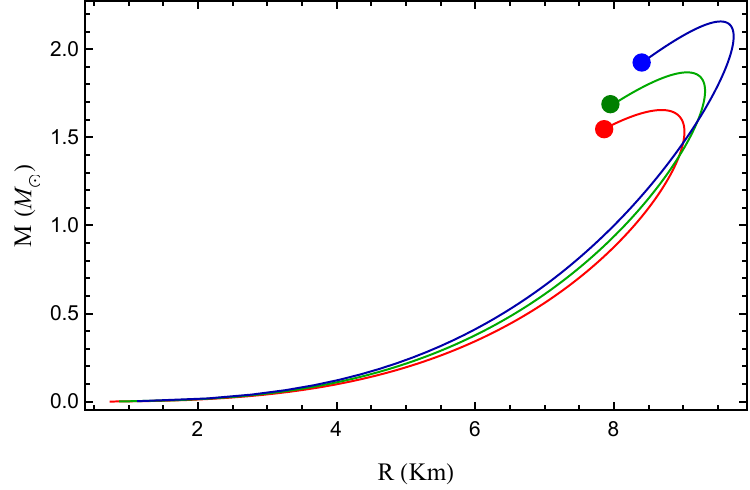}
			\subcaption{}
			\label{fig7}
		\end{minipage}
		\hfill
		\begin{minipage}{0.45\linewidth}
			\hspace{-0.4cm}
			\includegraphics[width=8.5cm]{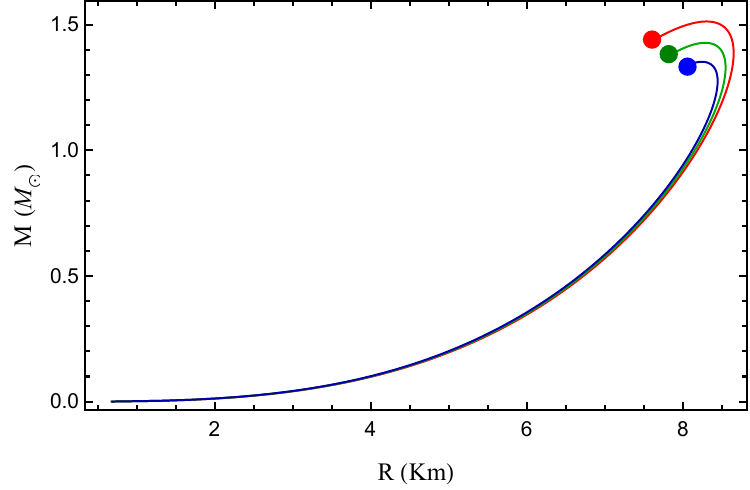}
			\subcaption{}
			\label{fig8}
		\end{minipage}
		\caption{Mass-radius relation for $B_{g}=95.11~\mathrm{MeV/fm^3}$, and different combinations of $\alpha$ and $\beta$. Figure~\ref{fig5}.: $\beta=-0.5$ with $\alpha=0,~2.5,~5$ (red, green, blue). Figure~\ref{fig6}.: $\beta=0$ with $\alpha=0,~2.5,~5$ (red, green, blue). Figure~\ref{fig7}.: $\beta=0$ with  $\alpha=10,~30,~50$ (red, green, blue). Figure~\ref{fig8}.: $\beta=0.5$ with $\alpha=0,~2.5,~5$ (red, green, blue).}
		\label{fig2a}
	\end{figure*}	
	Additionally, the maximum mass and the associated radius corresponding to the Figs.~\ref{fig5}, \ref{fig6}, \ref{fig7}, and \ref{fig8} are tabulated in Tables~\ref{tab5}, \ref{tab6}, \ref{tab7}, and \ref{tab8}, respectively.
	
	\begin{table*}[t!]
		\centering
		\caption{Maximum mass and radius for $B_{g}=95.11~\mathrm{MeV/fm^3}$ with different values of $\alpha$ and $\beta$. 
			Panels (a)--(d) correspond to Figs.~\ref{fig5}--\ref{fig8}.}
		\vspace{0.3em}
		\begin{subtable}[t]{0.45\textwidth}
			\centering
			\caption{$\beta=-0.5$}
			\label{tab5}
			\begin{tabular}{ccc} \hline
				$\alpha$ & $M_{\rm max}(M_{\odot})$ & $R$ (km) \\ \hline
				0 & 1.62 & 8.78 \\
				2.5 & 1.76 & 8.84 \\
				5 & 1.90 & 8.88 \\ \hline
			\end{tabular}
		\end{subtable}
		\hspace{1em}
		\begin{subtable}[t]{0.45\textwidth}
			\centering
			\caption{$\beta=0$}
			\label{tab6}
			\begin{tabular}{ccc} \hline
				$\alpha$ & $M_{\rm max}(M_{\odot})$ & $R$ (km) \\ \hline
				0 & 1.56 & 8.53 \\
				2.5 & 1.58 & 8.57 \\
				5 & 1.61 & 8.61 \\ \hline
			\end{tabular}
		\end{subtable}
		\vspace{1cm}		
		\begin{subtable}[t]{0.45\textwidth}
			\centering
			\vspace{0.2cm}
			\caption{$\beta=0$ (large $\alpha$)}
			\label{tab7}
			\begin{tabular}{ccc} \hline
				$\alpha$ & $M_{\rm max}(M_{\odot})$ & $R$ (km) \\ \hline
				10 & 1.65 & 8.69 \\
				30 & 1.87 & 9.05 \\
				50 & 2.16 & 9.54 \\ \hline
			\end{tabular}
		\end{subtable}
		\hspace{1em}
		\begin{subtable}[t]{0.45\textwidth}
			\centering
			\vspace{0.2cm}
			\caption{$\beta=0.5$}
			\label{tab8}
			\begin{tabular}{ccc} \hline
				$\alpha$ & $M_{\mathrm{max}}(M_{\odot})$ & $R$ (km) \\ \hline
				0 & 1.51 & 8.30 \\
				2.5 & 1.43 & 8.28 \\
				5 & 1.35 & 8.24 \\ \hline
			\end{tabular}
		\end{subtable}
	\end{table*}
	We have noted the following results:
	\begin{itemize}
		\item As the bag constant increases from $57.55~\mathrm{MeV/fm^3}$ to $95.11~\mathrm{MeV/fm^3}$, the disparity between the perturbative and non-perturbative vacua becomes more pronounced. According to the MIT EoS, $p=\frac{1}{3}(\rho-4B_{g})$, a higher value of $B_{g}$ leads to reduced pressure support. These two factors contribute in softening the EoS, resulting in a lower maximum mass and radius, which is clearly observed from Tables~\ref{tab5}, \ref{tab6}, \ref{tab7}, and \ref{tab8}, when compared with the previous case. 
		\item In this regime of constant bag, the maximum mass increases with increasing $\alpha$ and a negative value of $\beta$. However, we have noted that for $\beta=-0.5$, regular and well-behaved TOV solutions can be obtained up to $\alpha=9.8$. Within that parameter range, the maximum mass and the corresponding radius are $2.21~\mathrm{M_{\odot}}$ and $8.85~\mathrm{Km}$. 
		\item For $\beta=0$, $\alpha$ can reach values as high as $78.20$, yielding a maximum mass of $3.11~\mathrm{M_{\odot}}$ and a radius of $11.16~\mathrm{Km}$. 
		\item For $\beta=0.5$, we have obtained the threshold value of $\alpha$ to be $\alpha=6.2$, producing a maximum mass of $1.32~\mathrm{M_{\odot}}$ and a radius of $8.21~\mathrm{Km}$. Acceptable solutions of TOV equations are forbidden for $\alpha>6.2$. Further, in the positive range of $\beta$, the gravity-matter coupling becomes strong, and the increasing curvature corrections reduce the effective pressure. As a result, the EoS becomes softer, and the maximum mass decreases. 
	\end{itemize}
\end{itemize}
\begin{itemize}
	\item {\bf Case-III:} Here, we have chosen particular values of $\alpha$, and appropriate variations of $\beta$ to obtain the maximum mass and radius. Following the previous cases, we have shown the results in Figs.~\ref{fig9}, \ref{fig10}, \ref{fig11} and \ref{fig12} for $B_{g}=57.55~\mathrm{MeV/fm^3}$ and $95.11~\mathrm{MeV/fm^3}$ respectively. 
	\begin{figure*}[t!]
		\centering
		\begin{minipage}{0.45\linewidth}
			\hspace{-0.5cm}
			\includegraphics[width=8.5cm]{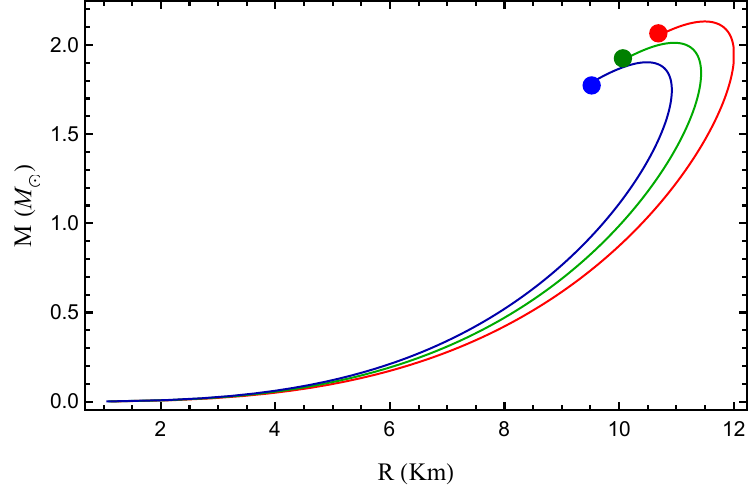}
			\subcaption{}
			\label{fig9}
		\end{minipage}
		\hfill
		\begin{minipage}{0.45\linewidth}
			\hspace{-0.4cm}
			\includegraphics[width=8.5cm]{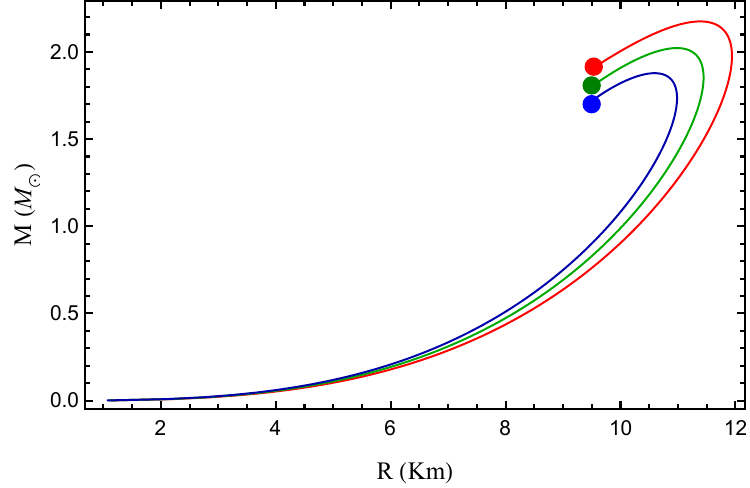}
			\subcaption{}
			\label{fig10}
		\end{minipage}
		\caption{Mass-radius relation for $B_{g}=57.55~\mathrm{MeV/fm^3}$ and different combinations of $\alpha$ and $\beta$. Figure~\ref{fig9}: $\alpha=0$, with $\beta=-0.82,~0$, and $0.82$ (red, green, blue). Figure~\ref{fig10}: $\alpha=1.5$ with $\beta=-0.61,~0$ and, $0.61$ (red, green, blue).}
	\end{figure*}
	
	\begin{table*}[t!]
		\centering
		\caption{Maximum mass and radius for $B_{g}=57.55~\mathrm{MeV/fm^3}$ with different $\alpha$ and $\beta$ values. Panels (a) and (b) correspond to Figs.~\ref{fig9} and \ref{fig10} respectively.}			
		\begin{subtable}[t]{0.45\textwidth}
			\centering
			\caption{$\alpha=0$}
			\label{tab9}
			\begin{tabular}{ccc} \hline
				$\beta$ & $M_{\rm max}(M_{\odot})$ & $R$ (Km) \\ \hline
				-0.82 & 2.13 & 11.50 \\
				0 & 2.012 & 10.96 \\
				0.82 & 1.90 & 10.49 \\ \hline
			\end{tabular}
		\end{subtable}
		\hspace{1em}
		\begin{subtable}[t]{0.45\textwidth}
			\centering
			\caption{$\alpha=1.5$}
			\label{tab10}
			\begin{tabular}{ccc} \hline
				$\beta$ & $M_{\rm max}(M_{\odot})$ & $R$ (Km) \\ \hline
				-0.61 & 2.17 & 11.39 \\
				0 & 2.02 & 10.98 \\
				0.61 & 1.87 & 10.59 \\ \hline
			\end{tabular}
		\end{subtable}
	\end{table*}
	\begin{figure*}
		\begin{minipage}{0.45\linewidth}
			\hspace{-0.5cm}
			\includegraphics[width=8.5cm]{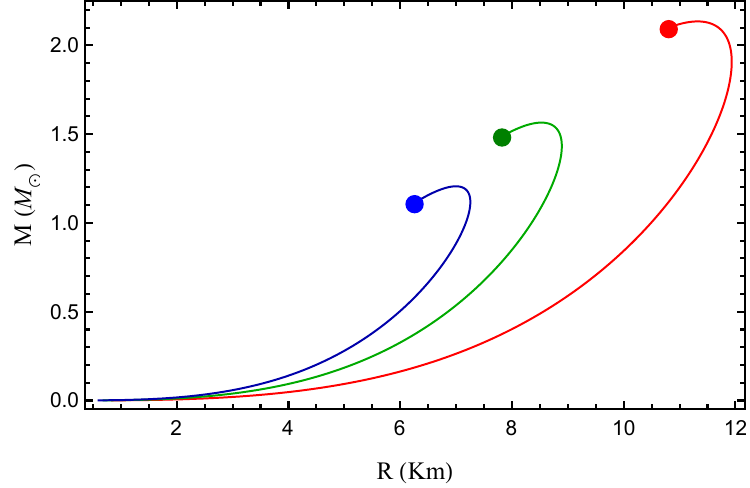}
			\subcaption{}
			\label{fig11}
		\end{minipage}
		\hfill
		\begin{minipage}{0.45\linewidth}
			\hspace{-0.4cm}
			\includegraphics[width=8.5cm]{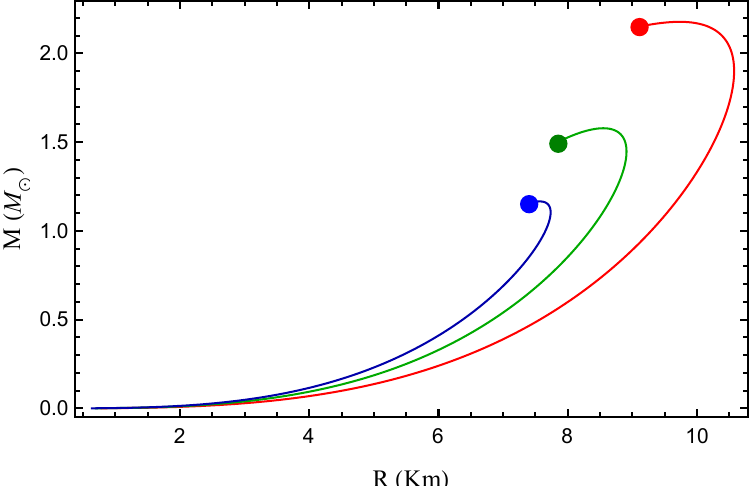}
			\subcaption{}
			\label{fig12}
		\end{minipage}
		\caption{Mass-radius relation for $B_{g}=95.11~\mathrm{MeV/fm^3}$  different values and different combinations of $\alpha$ and $\beta$. Figure~\ref{fig11}: $\alpha=0$, with $\beta=-4.1,~0$, and $4.1$ (red, green, blue). Figure~\ref{fig10}: $\alpha=1.5$ with $\beta=-2.2,~0$, and $1.9$ (red, green, blue).}
	\end{figure*}
	Furthermore, we have extracted the maximum mass-radius results from the plots and they are tabulated in Tables~\ref{tab9}, \ref{tab10}, \ref{tab11}, and \ref{tab12}.
	
	\begin{table*}[t!]
		\centering
		\caption{Tabulation of maximum mass and radius for $B_{g}=95.11~\mathrm{MeV/fm^3}$, and different values of $\alpha$ and $\beta$. Panels (a) and (b) correspond to Figs.~\ref{fig11} and \ref{fig12} respectively.}
		\begin{subtable}[t]{0.45\textwidth}
			\centering
			\caption{$\alpha=0$}
			\label{tab11}
			\begin{tabular}{ccc}\hline
				$\beta$ & $M_{\rm max}(M_{\odot})$ & $R$ (Km) \\ \hline
				-4.1 & 2.13 & 11.32 \\
				0    & 1.56 & 8.53  \\
				4.1  & 1.21 & 7.00  \\ \hline
			\end{tabular}
		\end{subtable}
		\hspace{1em}
		\begin{subtable}[t]{0.45\textwidth}
			\centering
			\caption{$\alpha=1.5$}
			\label{tab12}
			\begin{tabular}{ccc}\hline
				$\beta$ & $M_{\rm max}(M_{\odot})$ & $R$ (Km) \\ \hline
				-2.2 & 2.17 & 9.73 \\
				0    & 1.58 & 8.55 \\
				1.9  & 1.16 & 7.57 \\\hline
			\end{tabular}
		\end{subtable}
	\end{table*}
	From Figs.~\ref{fig9}, \ref{fig10}, \ref{fig11}, and \ref{fig12}, we infer the following:
	\begin{itemize}
		\item A lower bag constant leads to a stiffer EoS, resulting in reduced compressibility of the stellar structure. As a result, both the maximum mass and the corresponding radius increase. A comparative analysis of Tables~\ref{tab9}, \ref{tab10} with Tables~\ref{tab11}, \ref{tab12} clearly shows that the maximum mass is greater for $B_{g}=57.55~\mathrm{MeV/fm^3}$ compared to $B_{g}=95.11~\mathrm{MeV/fm^3}$.
		\item With increasing gravity-matter coupling, the maximum mass decreases in all cases. 
		\item From Table~\ref{tab9}, it must be noted that for $B_{g}=57.55\\\mathrm{MeV/fm^3}$ and $\alpha=0$, the negative threshold of $\beta$ is -0.82 where as the positive range can increase up to $\beta=23$, yielding, a maximum mass of $0.67~\mathrm{M_{\odot}}$ and a radius of $5.22~Km$. Beyond this entire range, the TOV equations do not produce usable results. 
		\item From Table~\ref{tab10}, it is observed that for $B_{g}=57.55\\\mathrm{MeV/fm^3}$ and $\alpha=1.5$, the parameter $\beta$ admits a viable range from the lower threshold of $\beta=-0.61$ up to a maximum of $\beta=3.2$. Within this interval, the TOV equations yield physically acceptable solutions, with the maximum mass reaching $1.35~\mathrm{M_{\odot}}$ and the corresponding radius being $9.11~Km$. Outside this range of $\beta$, the TOV solutions become non-physical.
		\item Similar arguments for the range of $\beta$ are also applicable for the Tables~\ref{tab11} and \ref{tab12} pertaining to $B_{g}=95.11~\mathrm{MeV/fm^3}$. In Table~\ref{tab11}, for $\alpha=0$, the maximum allowed value is $\beta=13$ with a maximum mass $0.77~\mathrm{M_{\odot}}$ and radius $5.17~Km$. On the other hand, from Table~\ref{tab12}, we note that $\beta$ does not converge to a symmetric limit, {\it viz.}, from -2.2 to 2.2, rather the maximum permissible value is $\beta=1.9$. 
	\end{itemize}
\end{itemize}
\section{Stability Analysis}\label{sec5} The validity of any newly proposed model relies on the stability analysis. In the present context, since we have used a particular barotropic EoS, we have studied the variation of adiabatic index to substantiate this aspect. Further, the proposed analysis revolves around the determination of maximum mass-radius in this novel framework. Consequently, we have studied the Harrison-Zel'dovich-Novikov criterion \cite{Zeldovich} to assess the stability of the model. Notably, we have selected the following range of parameters in for this analysis: $B_{g}=57.55$ and $95.11~\mathrm{MeV/fm^{3}}$, $\beta=0.5$, and $\alpha=0,~2.5$ and $5$.
\subsection{Radial variation of adiabatic index} 
\begin{figure*}[t]
	\begin{minipage}{0.45\linewidth}
		\hspace{-0.5cm}
		\includegraphics[width=8.5cm]{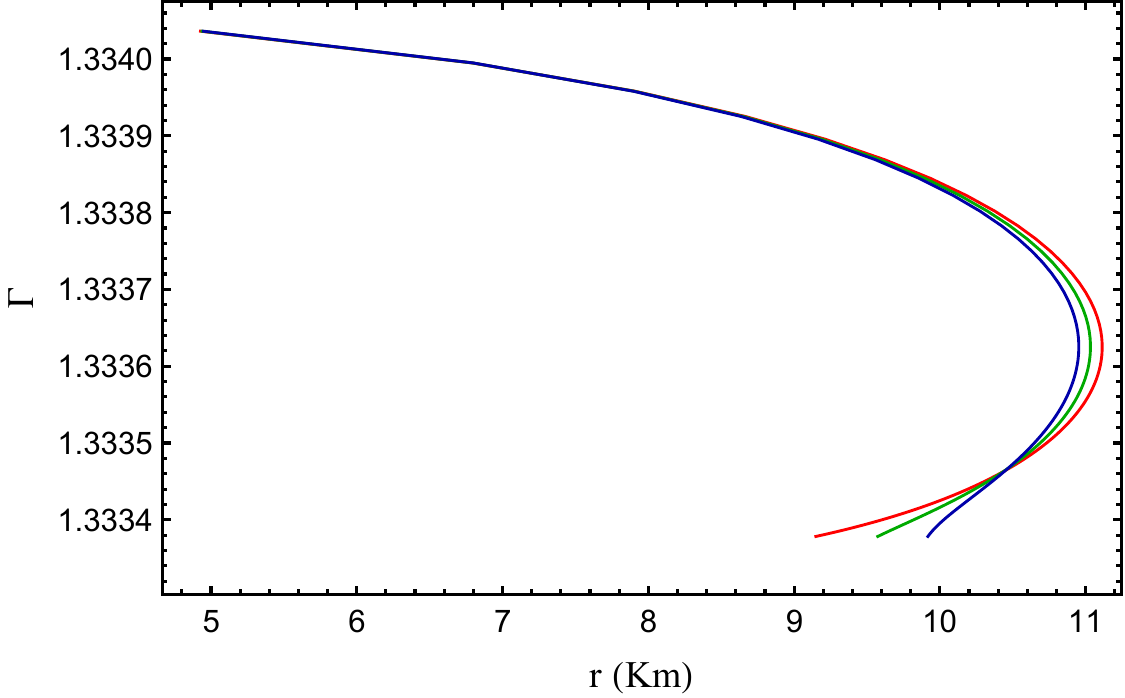}
		\subcaption{}
		\label{fig13}
	\end{minipage}
	\hfill
	\begin{minipage}{0.45\linewidth}
		\hspace{-0.4cm}
		\includegraphics[width=8.5cm]{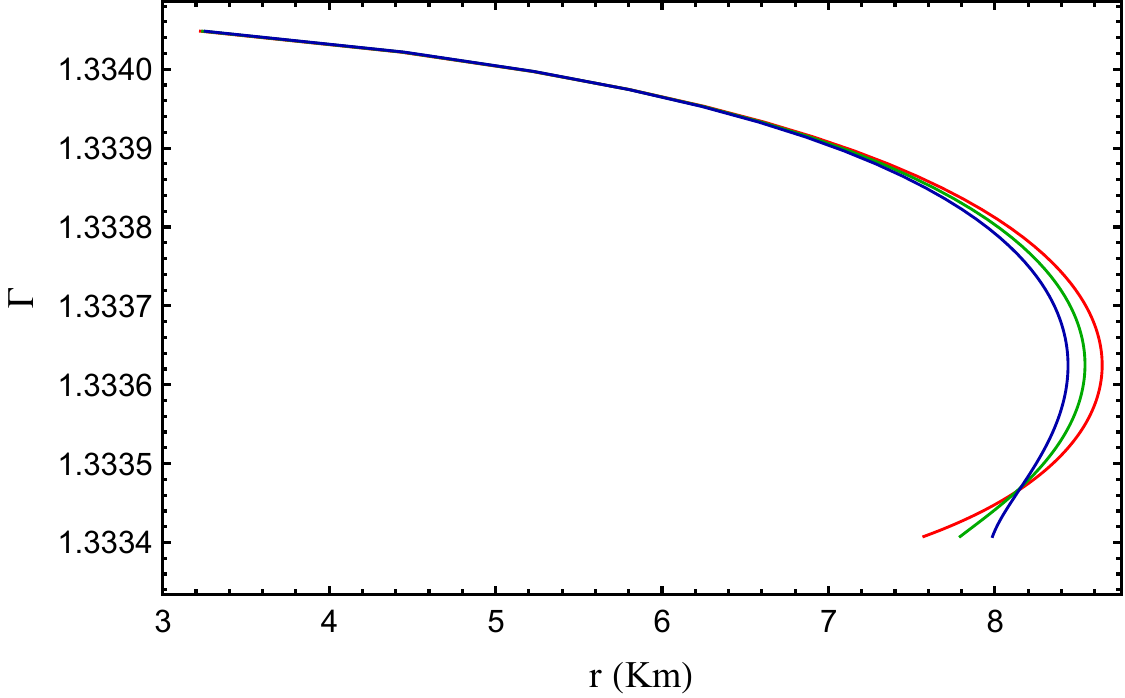}
		\subcaption{}
		\label{fig14}
	\end{minipage}
	\caption{Radial variation of adiabatic index for (a) $B_{g}=57.55~\mathrm{MeV/fm^3}$ and (b) $B_{g}=95.11~\mathrm{MeV/fm^3}$. Here, $\beta=0.5$ and red, green, blue lines represent $\alpha=0,~2.5,$ and $5$, respectively.}
	\label{fig5a}
\end{figure*}

To validate the use of a barotropic EoS, such as the MIT bag model EoS, within this proposed theory of gravity, we analyse the adiabatic index within the present parameter space. Adiabatic index $(\Gamma)$ represents the ratio of specific heat that embodies the stiffness of an equation of state. For an isotropic stellar configuration, the adiabatic index is written in the following form:
\begin{equation}
	\Gamma=\left(1+\frac{\rho}{p}\right)\left(\frac{dp}{d\rho}\right), \label{eq12}
\end{equation} 
where $\left(\frac{dp}{d\rho}\right)$ represents the square of the sound velocity. Notably, we have used the MIT bag model EoS here, hence, $\left(\frac{dp}{d\rho}\right)=\frac{1}{3}$. Now, in an isotropic matter configuration, Heintzmann and Hillebrandt \cite{Heintzmann} showed that in the framework of macroscopic stability analysis for compact star models subjected to small radial perturbations, maintaining a real eigenfrequency for the fundamental radial oscillation mode requires the adiabatic index to be greater than $\frac{4}{3}$. Now, considering the present framework, we have graphically illustrated the radial variation of adiabatic index in Fig.~\ref{fig5a}. It must be noted that we have obtained these variations from the solution of TOV equations coupled with MIT bag model EoS. 
From Fig.~\ref{fig5a}, we note that the condition $\Gamma>\frac{4}{3}\approx1.33$ is well-satisfied in this framework, which further substantiates our proposed framework.

\subsection{Harrison-Zel'dovich-Novikov criterion} 
\begin{figure*}[t]
	\begin{minipage}{0.45\linewidth}
		\hspace{-0.5cm}
		\includegraphics[width=8.5cm]{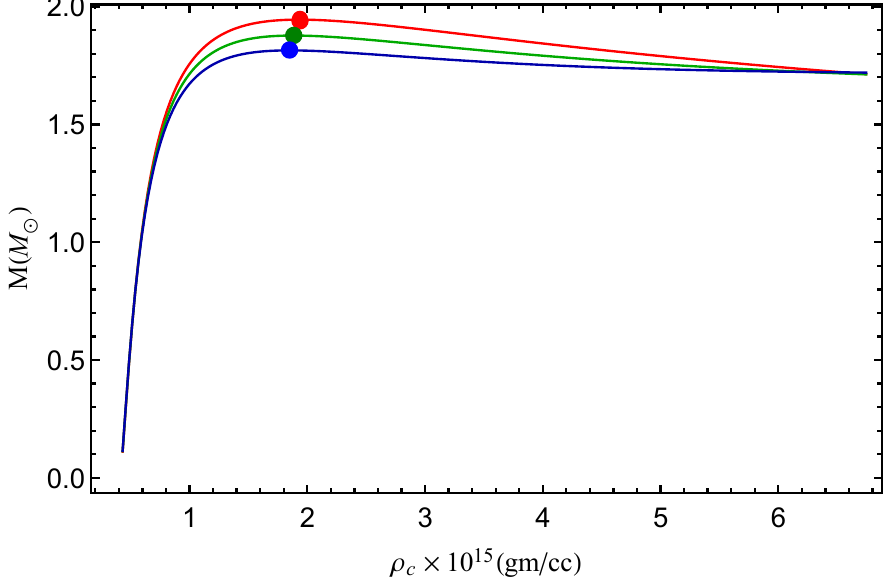}
		\subcaption{}
		\label{fig15}
	\end{minipage}
	\hfill
	\begin{minipage}{0.45\linewidth}
		\hspace{-0.4cm}
		\includegraphics[width=8.5cm]{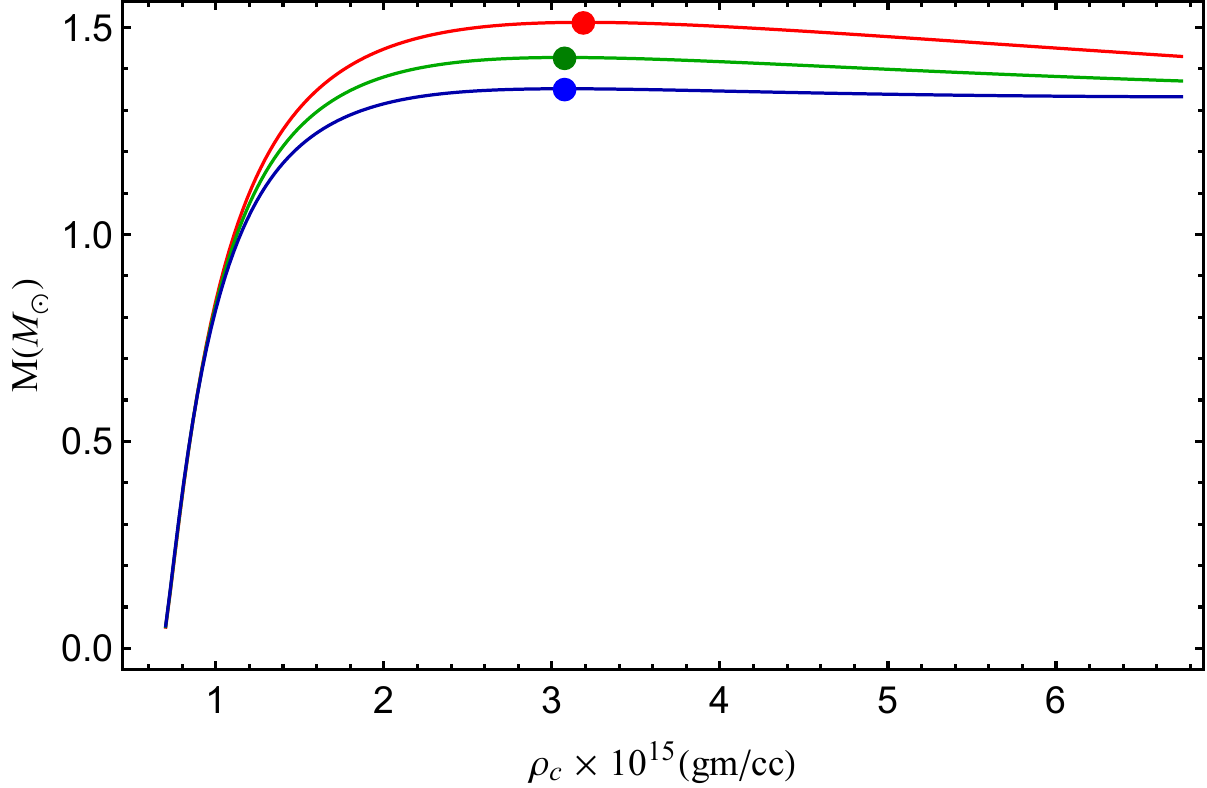}
		\subcaption{}
		\label{fig16}
	\end{minipage}
	\caption{Mass-central density for (a) $B_{g}=57.55~\mathrm{MeV/fm^3}$ and (b) $B_{g}=95.11~\mathrm{MeV/fm^3}$. Here, $\beta=0.5$ and red, green, blue lines represent $\alpha=0,~2.5,$ and $5$, respectively.}
	\label{fig6a}
\end{figure*}

\begin{figure*}[t]
	\begin{minipage}{0.45\linewidth}
		\hspace{-0.5cm}
		\includegraphics[width=8.5cm]{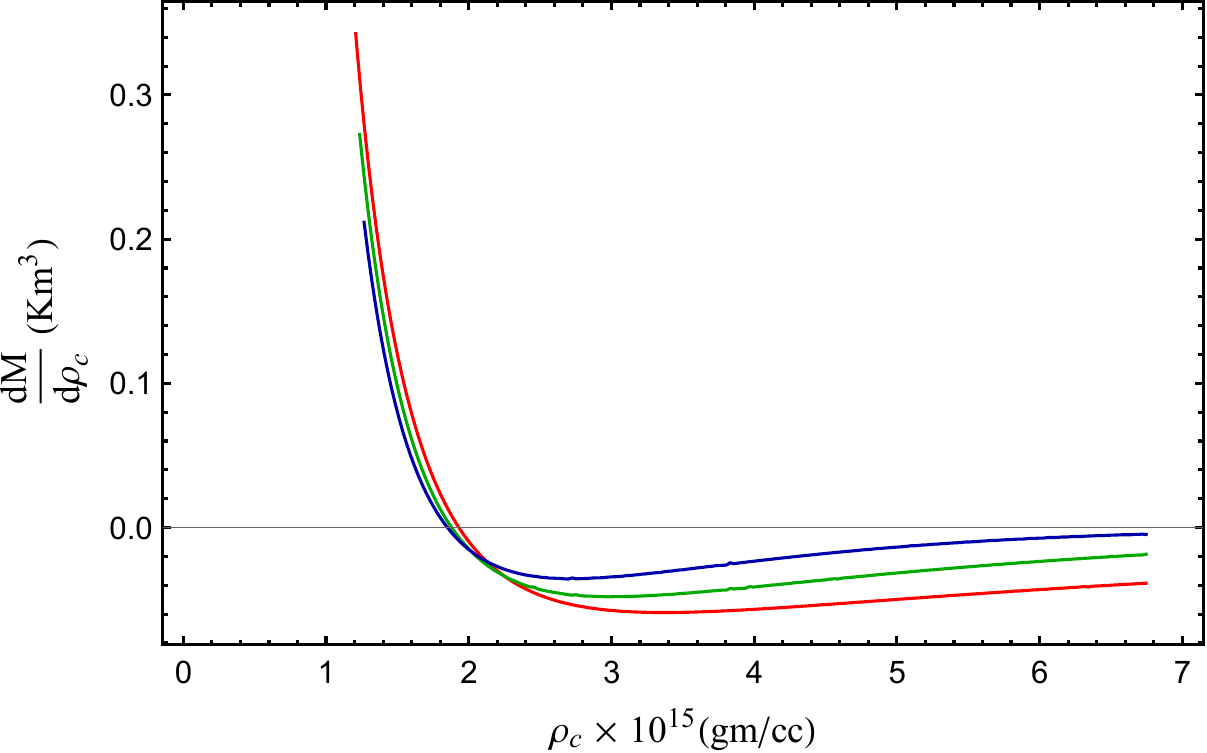}
		\subcaption{}
		\label{fig17}
	\end{minipage}
	\hfill
	\begin{minipage}{0.45\linewidth}
		\hspace{-0.4cm}
		\includegraphics[width=8.5cm]{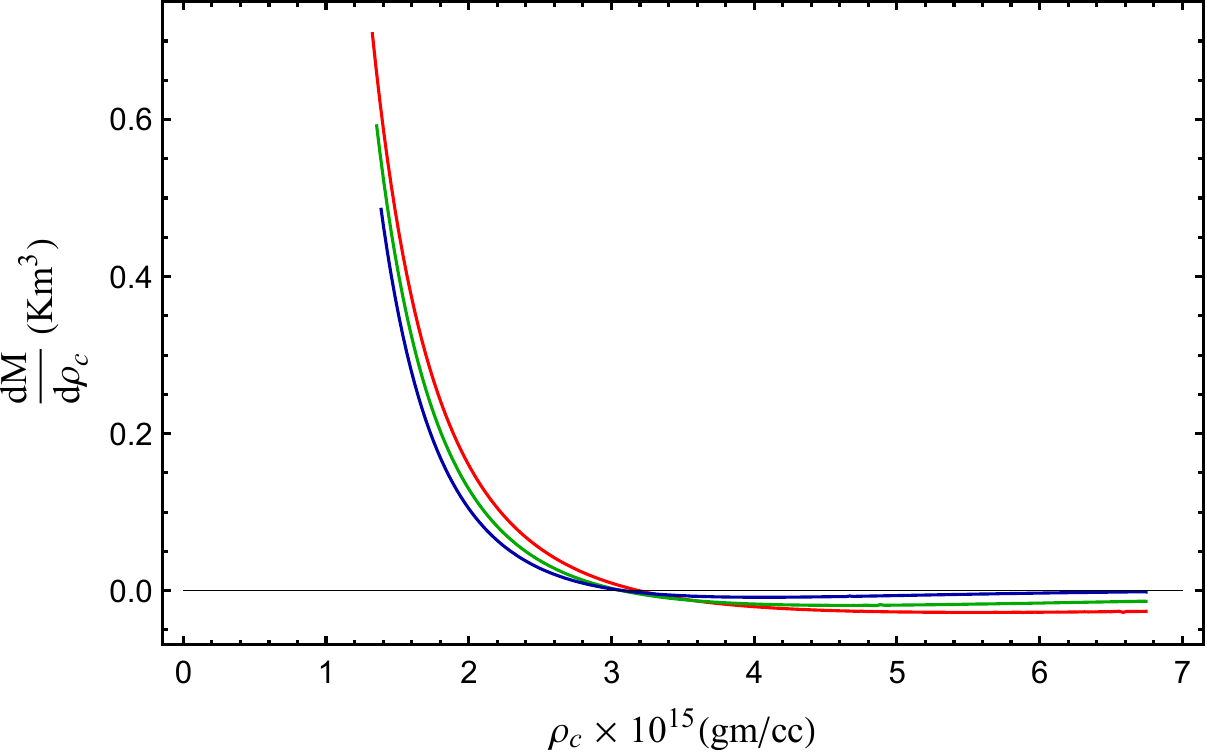}
		\subcaption{}
		\label{fig18}
	\end{minipage}
	\caption{$\frac{dM}{d\rho_{c}}$ vs $\rho_{c}$ for (a) $B_{g}=57.55~\mathrm{MeV/fm^3}$ and (b) $B_{g}=95.11~\mathrm{MeV/fm^3}$. Here, $\beta=0.5$ and red, green, blue lines represent $\alpha=0,~2.5,$ and $5$, respectively.}
	\label{fig7a}
\end{figure*}

In literature, the Harrison-Zel'dovich-Novikov criterion \cite{Zeldovich} is a standard analysis pertaining to the stability of stellar configuration and characterised by the computation of $\Big(\frac{dM}{d\rho_{c}}\Big)$, where $M$ is the maximum mass and $\rho_{c}$ is the central density of the stellar structure. The solution of TOV equations has provided central densities that lead up to the maximum mass. Any further increases in the central density will increase the maximum mass thereby increasing the gravitational pull leading to collapse. This feature is accurately described by the graphical representations in Figs.~\ref{fig6a} and \ref{fig7a}. In Fig.~\ref{fig6a}, we have plotted the mass vs central density, where the maximum mass points are denoted by the solid dots. Next, we have plotted $\Big(\frac{dM}{d\rho_{c}}\Big)$ vs $\rho_{c}$ in Fig.~\ref{fig7a}.     
From Fig.~\ref{fig7a}, we note that the maximum mass points and the corresponding central density describe the onset of instability. Beyond these points $\Big(\frac{dM}{d\rho_{c}}\Big)<0$ describing a collapsed or unstable configuration. To solidify this aspect, we have tabulated the parameters at the turning point in Table~\ref{tab12a}. 
\begin{table*}[t!]
	\centering
	\caption{Parameters obtained at the turning point.}
	\begin{tabular}{lcccc}
		\hline
		\multirow{2}{*}{$B_{g}~(\mathrm{MeV/fm^{3}})$} & \multirow{2}{*}{$\beta$} & \multirow{2}{*}{$\alpha$} & \multicolumn{2}{c}{At the turning point}\vspace{0.1cm}\\
		& & & $\rho_{c}~(\times10^{15}~\mathrm{gm/cc})$ & $\frac{dM}{d\rho_{c}}$ \vspace{0.1cm}\\	\hline
		\multirow{3}{*}{57.55} & \multirow{3}{*}{0.5} & 0 & 1.93 & 8.89$\times10^{-18}$ \\ 
		&& 2.5 & 1.88 & 6.05$\times10^{-17}$ \\
		&& 5 & 1.85 & 2.12$\times10^{-16}$ \\  
		\hline
		\multirow{3}{*}{95.11} & \multirow{3}{*}{0.5} & 0 & 3.19 & 7.59$\times10^{-19}$ \\ 
		&& 2.5 & 3.07 & 1.54$\times10^{-17}$ \\
		&& 5 & 3.08 & 3.94$\times10^{-17}$ \\   \hline
	\end{tabular}
	\label{tab12a}
\end{table*}
\section{Detectability of strange stars and constraining the parameter space}\label{sec6}
The present $f(\tilde{R},T)=R+\alpha R^{2}+2\beta T$ framework leads to several distinctive astrophysical signatures that enable both the detectability of strange stars and direct constraints on the model parameter space. A key outcome of our analysis is the significant enhancement of the maximum stable mass, reaching the value as high as $\sim3.11~M_\odot$. Such large masses fall within the so-called mass gap between canonical neutron stars and low-mass black holes, thereby opening the possibility that some compact objects detected in GW events, most notably the secondary component of GW190814, may be strange stars rather than black holes. In particular, a compact object with mass $M\gtrsim2.5~M_\odot$ and radius of order $\sim10$~Km would be difficult to reconcile with standard hadronic equations of state in GR, while being naturally accommodated within the present model.

The increased compactness of strange star configurations in this theory also leaves a clear imprint on tidal interactions during binary inspirals. Specifically, the tidal Love numbers and the associated tidal deformability $\Lambda(M)$ are systematically reduced relative to GR predictions. Consequently, joint constraints on masses and tidal deformabilities from LIGO-Virgo-KAGRA observations provide a powerful and quantitative test of the allowed $(\alpha,\beta)$ parameter space. Future high-precision measurements of inspiral waveforms will be particularly effective in discriminating between modified-gravity strange star models and standard NSs.

Beyond the inspiral phase, the raised maximum mass significantly affects post-merger dynamics. The delayed collapse of merger remnants implies the formation of longer-lived hypermassive or supramassive configurations, leading to characteristic shifts in the dominant post-merger gravitational wave frequencies. Targeted searches for such spectral features in high signal-to-noise ratio detections may therefore provide an additional avenue to distinguish the present framework from GR-based expectations.

Electromagnetic observations further complement gravitational wave probes. Strange stars, especially those with partially or fully bare quark surfaces, are expected to exhibit rapid neutrino cooling and distinctive thermal emission properties. These effects translate into characteristic X-ray cooling curves and spectral energy distributions that differ from those of hadronic NSs. Current and upcoming X-ray missions, such as NICER and XMM-Newton, can thus place meaningful constraints on the viability of the model.

Another promising observation lies in the moment of inertia and its relation to other macroscopic stellar properties. Owing to modifications in the internal structure induced by curvature-matter coupling, deviations from the universal I-Love-Q relations of GR are anticipated. Precise measurements of the moment of inertia in double pulsar systems would therefore offer a stringent test of the theory and help narrow down the admissible values of $\alpha$ and $\beta$.

Finally, pulsar timing observations provide complementary information through glitch activity. The presence or absence of a conventional crust in strange stars influences glitch amplitudes and recurrence statistics. Correlating glitch behaviour with independently inferred stellar masses and radii could offer indirect evidence in favour of-or against-the strange star interpretation within this modified gravity framework.

A multi-messenger observational strategy combining mass-radius measurements (NICER), tidal deformability constraints (LIGO/Virgo/KAGRA), moment of inertia determinations (radio timing), and thermal emission studies (NICER/XMM-Newton) enables a comprehensive mapping of the allowed $(\alpha,\beta)$ parameter space. Such approaches also ensures consistency with existing solar-system and post-Newtonian bounds, thereby providing a robust pathway to validate or rule out the present theory.

\section{Conclusions}\label{sec7} 
In the present paper, we have re-examined the maximum mass limit of strange stars in the context of $f(\tilde{R},T)=R+\alpha R^{2}+2\beta T$ gravity. Considering a static, spherically symmetric space-time and an isotropic perfect fluid distribution, we have reformulated the TOV equations in this framework and they are expressed in Eqs.~(\ref{eq10}) and (\ref{eq11}). Since, the solutions of the TOV equations require an EoS, we have considered the prevalent MIT bag model EoS, $p=\frac{1}{3}(\rho-4B_{g})$, where, $B_{g}$ is the bag parameter. Following the work of Refs. \cite{Kettner,Madsen}, it is well-established that in the approximation of massless quarks, the range of bag parameter $(B_{g})$ is $57.55-95.11~\mathrm{MeV/fm^3}$ that ensures stable quark matter configuration. Now, the resulting M-R plots are subjected to a parameter space that is constrained by the choice of $\alpha$, $\beta$ and $B_{g}$. To start with, we have considered the positive values of $\alpha$ to discard any ghost fields that may appear, and building on the earlier investigations \cite{Deb,Carvalho}, we have observed that the parameter $\beta$ may assume both positive and negative values. Accordingly, we have systematically varied both $\alpha$ and $\beta$ to explore their influence on the maximum mass-radius relationship. The parameter ranges for $\alpha$ and $\beta$ are selected judiciously to ensure that the resulting solutions to the TOV equations remain physically consistent and mathematically stable. The quadratic curvature term $\alpha R^{2}$ introduces a scalar degree of freedom whose viability requires $f_{RR}>0$, implying $\alpha>0$, and a sufficiently large scalaron mass in the weak-field regime. These conditions ensure consistency with Solar-System and post-Newtonian tests, while allowing deviations from GR to become relevant only in high-curvature environments such as compact star interiors. The values of $\alpha$ adopted in this work are chosen within this theoretically motivated regime. The parameter $\beta$ governs the non-minimal matter-geometry coupling through the linear $T$ term. This coupling does not introduce an additional propagating degree of freedom but leads to controlled deviations from energy-momentum conservation. Existing weak-field constraints on $\beta$ are highly model dependent. Therefore, we restrict $|\beta|$ to small values so that deviations from GR remain perturbative and compatible with post-Newtonian bounds. From a cosmological standpoint, the $R^{2}$ sector is well known to be consistent with inflationary and late-time cosmology, while the linear $T$ coupling primarily affects high-density matter configurations and does not significantly alter vacuum or radiation-dominated evolution. Hence, the explored $(\alpha,\beta)$ ranges are compatible with current post-Newtonian and cosmological constraints, while being tailored to probe strong-field effects inside compact stars. Notably, in the limit $\alpha=0$ and $\beta=0$ the value of GR can be achieved from the present framework.

Based on the above specifications, we have noted that the negative values of $\beta$ correspond to a weakly coupled regime, which facilitates higher mass-radius configurations. This behaviour is observed consistently for both the values of the bag constant, $B_{g}=57.55~\mathrm{MeV/fm^3}$ and $B_{g}=95.11~\mathrm{MeV/fm^3}$. In this scenario, increasing $\alpha$ makes the EoS stiffer, which leads to increased maximum mass limits of SS as evident from Table~\ref{tab1}. Further, from Tables~\ref{tab2} and \ref{tab3}, we note that, for $\beta=0$, we are effectively working in the regime of $f(R)=R+\alpha R^{2}$ gravity and the maximum mass limit also increases here with increasing curvature corrections. On the contrary, the positive range of $\beta$ ensures a strong gravity-matter coupling which dominates over the higher curvature. As a result, with increasing $\alpha$, the maximum mass decreases, as shown in Table~\ref{tab4}. In case of $\alpha=0$ and $\beta=0$, this formalism yields a maximum mass of $2.012~\mathrm{M_{\odot}}$, that matches with the associated result for SS in GR. For $B_{g}=95.11~\mathrm{MeV/fm^3}$, the maximum mass increases with increasing $\alpha$ for $\beta\leq0$. However, the mass limit is lower in comparison to the case of $B_{g}=57.55~\mathrm{MeV/fm^3}$, as described in Tables~\ref{tab5}, \ref{tab6} and \ref{tab7}. This is because, even though increasing $\alpha$ and considering $\beta\leq0$ may result in a stiffer EoS, the increased $B_{g}$ value increases the difference between perturbative and non-perturbative vacuum, which in turn leads to a more confined configuration. However, from Table~\ref{tab8}, it is noted that with $\beta>0$, the maximum mass decreases with increasing $\alpha$. For $\alpha\geq0$, the variation of $\beta$ yields that the maximum mass decreases with increasing $\beta$ and the results are tabulated in Tables~\ref{tab9}, \ref{tab10}, \ref{tab11} and \ref{tab12}. Moreover, this nature is same for $B_{g}=57.55~\mathrm{MeV/fm^3}$ and $B_{g}=95.11~\mathrm{MeV/fm^3}$. Through this novel formalism, we have obtained a new extended maximum mass limit of SS in modified gravity, and notably under different parametric choices, the conventional maximum mass limit of SS is increased. 

In this model, an interesting correlation exists between $\alpha$ and $\beta$. For a particular bag parameter, if $\alpha$ is increased, the range of $\beta$ becomes narrow. When the bag parameter is increased, the corresponding numerical values of $\beta$ increase, but the same characteristic narrowing of the $\beta$ range persists with increasing $\alpha$. In the present framework, it is interesting to note that for $B_{g}=57.55~\mathrm{MeV/fm^3}$ the value of gravity-matter coupling $\beta\rightarrow0$, if we consider the value of $\alpha>52$. Same argument is valid for $B_{g}=95.11~\mathrm{MeV/fm^3}$ with $\beta\rightarrow0$ for $\alpha>78$. Therefore, if such high curvature corrections are involved, the gravity-matter coupling becomes insignificant. However, under such criteria, the present formalism can achieve a maximum mass of $2.42~\mathrm{M_{\odot}}$ for $\beta=0$, $\alpha=52$ and $B_{g}=57.55~\mathrm{MeV/fm^3}$ and $3.11~\mathrm{M_{\odot}}$ for $\beta=0$, $\alpha=78.20$ and $B_{g}=95.11~\mathrm{MeV/fm^3}$. Hence, we may elucidate that with increasing bag parameter, the range of $\alpha$ increases in the framework of $f(R+\alpha R^{2})$ gravity, which significantly increases the maximum mass in this scenario. Based on this result, it may be inferred that the inclusion of higher order curvature and non-minimal matter coupling provides an extended framework, within which the lighter companion of GW190814 with a maximum mass of $2.59^{+0.08}_{-0.09}~\mathrm{M_{\odot}}$ \cite{Abbott}, may also be interpreted as a strange star. 

The recent observations of Low-Mass X-ray Binaries using Rossi X-Ray Timing Explorer (RXTE) and XMM-Newton have provided the mass-radius of the NSs present in those binary systems. In addition, improvements in astronomical instrumentation have allowed us to constrain the mass-radius ranges of the lighter companions involved in GW events. To assess the validity of our theoretical model, we computed the radii of these compact stars and of the secondary objects in GW-events by solving the TOV equations within our current framework. The resulting predictions are summarised in Table~\ref{tab12aa}:
\begin{table*}[t]
	\centering
	\caption{Radius prediction of recently observed NSs and GW events.}
	\begin{tabular}{ccccccc}
		\hline
		Compact objects & Measured mass & Measured radius & $B_{g}$ & \multirow{2}{*}{$\beta$} & \multirow{2}{*}{$\alpha$} & Predicted radius \\
		& from observations $\mathrm{(M_{\odot})}$ & from observations $\mathrm{(Km)}$ & $(\mathrm{MeV/fm^{3}})$ & & & $\mathrm{(Km)}$ \\
		\hline
		\vspace{0.1cm}
		HER X-1 \cite{Abubekerov} & $0.85\pm0.15$ & $8.1\pm0.41$ & 85 & 0.5 & 5 & 8.1 \\
		\vspace{0.1cm}		
		EXO 1745-248 \cite{Ozel} & 1.4 & 11 & 60 & -0.45 & 4.4 & 11 \\
		\vspace{0.1cm}
		4U 1820-30 \cite{Guver} & $1.58\pm0.06$ & $9.1\pm0.4$ & 95.11 & -0.5 & 0.15 & 9.1 \\
		\vspace{0.1cm}
		4U 1608-52 \cite{Guver1} & $1.74\pm0.14$ & $9.3\pm1.0$ & 95.11 & 0.05 & 43 & 9.3 \\
		\vspace{0.1cm}
		GW170817 \cite{Abbott1} & 1.16-1.60 & $10.7^{+2.1}_{-1.5}$ & 57.55 & 0.3 & 1.1 & $10.7^{+0.39}_{-0.38}$ \\
		\vspace{0.1cm}
		GW190814 \cite{Abbott} & $2.59^{+0.08}_{-0.09}$ & -- & 95.11 & 0.02 & 72 & 10.41 \\ \hline
	\end{tabular}
	\label{tab12aa}
\end{table*}
From Table~\ref{tab12aa}, we note that the results obtained using the MIT bag model EoS demonstrate a notable agreement between theoretical predictions from the present model and estimation of radii of many compact stars including secondary objects of GW events from recent observations. This alignment suggests that, notwithstanding its simplified and phenomenological character, the EoS provides a realistic representation of the matter composition within compact stars (especially SS). Consequently, the present analysis supports the view that such a simplified framework may yield accurate and astrophysically relevant predictions, reinforcing the robustness of the conclusions drawn in this study.  

Notably, compared to previously established gravitational theories, the present formalism offers greater flexibility in exploring the maximum mass-radius range of compact objects. This extended range arises primarily from the inclusion of higher-order curvature corrections and the non-minimal matter-geometry coupling. We have presented a comparative analysis among different gravitational theories in terms of mass, radius and Buchdahl limits for compact stars in Table~\ref{tab13}.  
\begin{table*}[t!]
	\centering
	\caption{Comparative analysis of maximum mass, radius, and Buchdahl limits for compact stars in various modified gravity frameworks.}
	\begin{tabular}{lccccc}
		\hline
		Theory & $M_\mathrm{{\max}}$ & $R$ & Compactness  & Buchdahl Limit & Reference \\
		& $(M_\odot)$ & (Km) & $(\frac{M}{R})$ & $(\frac{M}{R}<\frac{4}{9})$ & \vspace{0.15cm}\\
		\hline
		GR (baseline) & $\approx2.0$ & $\approx11.0$ & 0.27 & Satisfied & P. Haensel et al. \cite{Haensel}  \\
		$f(R)$ & 1.75--2.07 & 9.1--9.9 & 0.28--0.31 & Satisfied & Astashenok et al. \cite{Astashenok} \\
		$f(R)=R+\alpha R^2$ & 2.05--2.51 & 9.94--11.30 & 0.30--0.33 & Satisfied & Feola et al. \cite{Feola} \\ 
		$f(R,T)$ & 2.07--2.60 & 8.87--11.24 & 0.34 & Satisfied & Deb et al. \cite{Deb1} \\
		$f(Q)$ & 1.95--2.65 & 14.96--15.64 & 0.19--0.25 & Satisfied & Lohakare et al. \cite{Lohakare} \\
		$f(Q,T)$ & 4.15 & 14.45 & 0.42 & Satisfied & Nashed and Harko \cite{Nashed}\\  
		$f(R+\alpha R^2,T)$ (this work) & 1.16--3.11 & 7.57--11.16 & 0.23--0.41 & Satisfied & Present study \\ \hline
	\end{tabular}
	\label{tab13}
\end{table*}
Furthermore, Figs.~\ref{fig5a}, \ref{fig6a}, and \ref{fig7a} confirm that the proposed framework satisfies the stability conditions, thereby strengthening the theoretical viability of the model. To reiterate, the entire formalism shows how the quadratic curvature with non-minimal matter coupling alters the maximum mass-radius limit of strange stars, and this feature places the work right in the intersection of modified gravity phenomenology and compact star astrophysics. The deviations from GR appearing in the analysis may open several potential avenues for new research. For example, as a possible research line, one may extend this framework to construct the tidal parameters, which could yield valuable insights into the tidal deformabilities of compact stars. With the rapid progress in gravitational wave observations, it may be possible to probe the tidal properties of binary mergers with unprecedented precision. Using this framework such parameters can be studied with a high degree of accuracy. Further, the wide mass-radius range may be useful to study the compact stars in a more precise manner using the NICER and radio timing constraints. Apart from these, the future research may delve into scrutinising the validity of energy conditions in this framework which is also a great way to test the viability of this study. In this regard, we have summarised the possibilities of SS detections and constraining the present parameter space in Table~\ref{tab14}.
\begin{table*}[t!]
	\centering
	\caption{Observational signatures of strange stars in 
		$f(\tilde R,T)=R+\alpha R^{2}+2\beta T$ gravity and corresponding constraints on the model.}
	\begin{tabular}{p{4.2cm} p{6.2cm} p{3.6cm}}
		\hline
		Observable & Predicted signature & Constrains \\
		\hline
		High-mass compact objects & Stable configurations up to $\sim3.11\,M_\odot$; mass-gap objects with $R\sim10$ km & $(\alpha,\beta)$ upper bounds \\
		Tidal deformability & Reduced $\Lambda(M)$ due to higher compactness & GW inspiral posteriors \\
		Post-merger GW spectrum & Delayed collapse; shifted dominant frequencies & Maximum mass, stiffness \\
		Thermal emission and cooling & Rapid neutrino cooling; distinct X-ray spectra & Surface composition, $\beta$ \\
		Moment of inertia & Deviations from GR I--Love--Q relations & Internal structure, $\alpha$ \\
		Pulsar glitches & Modified glitch statistics due to crust properties & Strange star viability \\
		Multi-messenger consistency & Joint GW, X-ray, and radio constraints & Allowed $(\alpha,\beta)$ region \\ \hline
	\end{tabular}
	\label{tab14}
\end{table*}

Finally, we may conclude that these findings demonstrate the potential of higher-order curvature-matter coupled gravity to accommodate heavier strange stars, offering a promising framework for probing the fundamental nature of dense matter and strong-field gravity in extreme astrophysical environments.
\section*{Acknowledgments}
DB is thankful to the Department of Science and Technology (DST), Govt. of India, for providing the fellowship vide no:  DST/INSPIRE Fellowship/2021/IF210761. PKC gratefully acknowledges support from IUCAA, Pune, India under Visiting Associateship programme. DB is also grateful to IUCAA, Pune, India for providing the visitor facilities. DB and PKC gratefully acknowledge the facilities provided by IUCAA during their visit, where this work was completed. The work of KB was partially supported by the JSPS KAKENHI Grant Numbers 24KF0100, 25KF0176 and Competitive Research Funds for Fukushima University Faculty (25RK011). We would like to express our gratitude to the respected referee for the meaningful comments and suggestions. 
\section{Declarations}
\begin{description}
	\item[Funding:] Not applicable. 
	\item[Conflicts of interest:] Not applicable 
	\item[Availability of data and material:] This manuscript has no associated data or the data will not be deposited, we have used only obtained mass-radius relations in the context of compact objects to construct relativistic stellar models.
\end{description}

\end{document}